\documentclass[a4paper, 10pt, conference]{ieeeconf}      
\usepackage{FG2024}

\FGfinalcopy 

\IEEEoverridecommandlockouts                              
\usepackage{graphics} 
\usepackage{epsfig} 
\usepackage{amsmath} 
\usepackage{amssymb}  
\usepackage{bm}
\usepackage{multirow}
\usepackage{hyperref}
\def\FGPaperID{****} 

\title{\LARGE \bf
Towards Bi-Hemispheric Emotion Mapping through EEG: A Dual-Stream Neural Network Approach}

\author{\parbox{16cm}{\centering
    {\large David Freire-Obregón, Daniel Hernández-Sosa, Oliverio J. Santana,}\\
    {\large Javier Lorenzo-Navarro, and Modesto Castrillón-Santana}\\
    {\normalsize
    SIANI, Universidad de Las Palmas de Gran Canaria, Las Palmas de Gran Canaria, Spain\\}}
    \thanks{This work is partially funded by the Spanish Ministry of Science and Innovation under project PID2021-122402OB-C22 and by the ACIISI-Gobierno de Canarias and European FEDER funds under project ULPGC Facilities Net and Grant \mbox{EIS 2021 04}.}
}

\begin{document}

\ifFGfinal
\thispagestyle{empty}
\pagestyle{empty}
\else
\author{Anonymous FG2024 submission\\ Paper ID \FGPaperID \\}
\pagestyle{plain}
\fi
\maketitle

\begin{abstract}

Emotion classification through EEG signals plays a significant role in psychology, neuroscience, and human-computer interaction. This paper addresses the challenge of mapping human emotions using EEG data in the Mapping Human Emotions through EEG Signals FG24 competition. Subjects mimic the facial expressions of an avatar, displaying fear, joy, anger, sadness, disgust, and surprise in a VR setting. EEG data is captured using a multi-channel sensor system to discern brain activity patterns. We propose a novel two-stream neural network employing a Bi-Hemispheric approach for emotion inference, surpassing baseline methods and enhancing emotion recognition accuracy. Additionally, we conduct a temporal analysis revealing that specific signal intervals at the beginning and end of the emotion stimulus sequence contribute significantly to improve accuracy. Leveraging insights gained from this temporal analysis, our approach offers enhanced performance in capturing subtle variations in the states of emotions. Code is available at \href{https://github.com/davidfreire/FG24-EmoNeuroDB/}{\textcolor{magenta}{https://github.com/davidfreire/FG24-EmoNeuroDB/}}

\end{abstract}

\begin{figure*}[t]  
    \centering
    \includegraphics[scale=0.60]{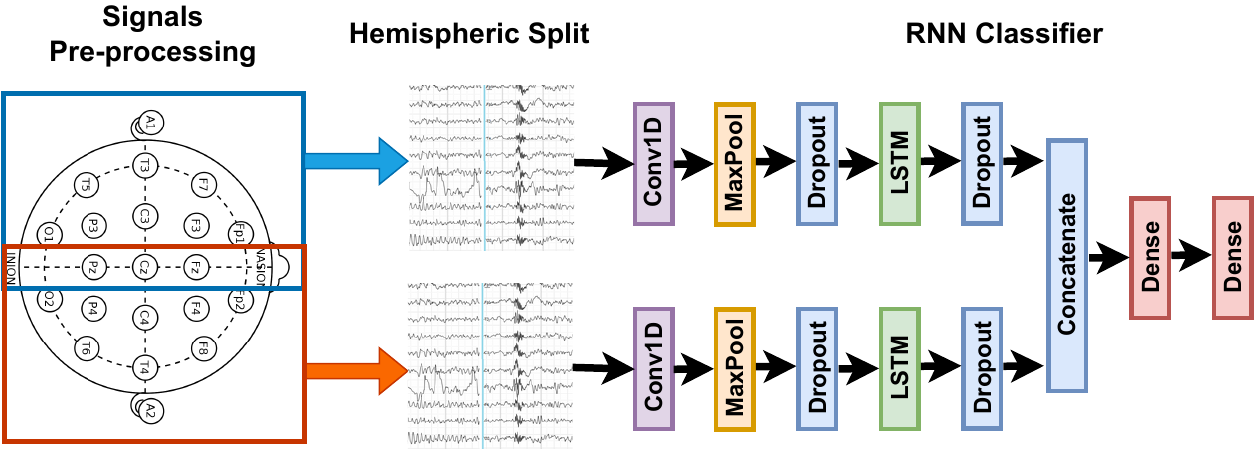}
    \caption{Diagram illustrating the three-module stages for Bi-Hemispheric emotion recognition from EEG signals: Signal Pre-processing, Hemispheric Split, and Signal Classification using RNN.}
     \label{fig:classifier}
\end{figure*}

\section{INTRODUCTION}

Emotion is a central aspect of human experience, profoundly shaping our interactions and perceptions. While humans naturally excel at discerning emotions in others, replicating this nuanced understanding in computers remains a formidable task \cite{Pantic03}. Emotion recognition serves as a crucial step towards imbuing machines with the ability to comprehend and respond to human emotions, garnering significant interest from researchers in human-machine interaction (HMI) and pattern recognition \cite{Freire09,Santana22}.

Traditionally, studies in emotion recognition have primarily focused on analyzing verbal and nonverbal cues, such as speech \cite{Mirsamadi17} and facial expressions \cite{Freire23}. However, recent insights from neuroscience suggest that emotions originate from various regions of the brain, including the orbital frontal cortex, ventral medial prefrontal cortex, and amygdala~\cite{Britton06}. This neurobiological perspective presents an intriguing opportunity to decode emotions by capturing continuous brain activity signals from these sub-cortical regions.

In the context of virtual reality (VR), where subjects are fully immersed in carefully crafted environments, understanding and recognizing emotions take on a new dimension. By utilizing electrodes placed on the scalp to perform an electroencephalogram (EEG), researchers can record neural activity within the brain while the subjects interact with virtual avatars displaying various emotion's expressions. This approach offers a unique opportunity to directly observe the neural mechanisms underlying the processing of emotions in a controlled VR environment.

Analyzing EEG signals within the VR paradigm provides valuable insights into how the brain responds to emotion's stimuli in immersive environments. Existing EEG-based emotion recognition techniques tackle two primary challenges: feature extraction and accurate classification. EEG signals offer rich data across time, frequency, and \mbox{time-frequency} domains, necessitating robust feature extraction methods. 

Previous works have already leveraged machine learning techniques evaluating EEG features, underscoring the complexity of this task \cite{Jenke14}. Subsequent challenges lie in effectively classifying these features.  In this regard, some works have proposed a group of sparse canonical correlation analysis for simultaneous EEG channel selection and emotion recognition \cite{Zheng17}, while others integrated brain activation patterns to enhance emotion's recognition performance \cite{Li19}.

While these methodologies have demonstrated promising results on specific EEG emotion's datasets, there remains a need for further exploration and refinement to achieve robust and generalized emotion recognition systems. The proposed methodology contributes to the field of EEG-based emotion recognition in several significant ways:

\begin{itemize}
\item Incorporates a Bi-Hemispheric approach within a \mbox{two-stream} recurrent neural network, leveraging the distinct processing characteristics of each hemisphere to enhance emotion inference accuracy.
\item Conducts comparative analyses with one-stream neural networks, demonstrating the superiority of the \mbox{Bi-Hemispheric} approach in capturing subtle nuances of emotions.
\item Conducts a temporal analysis of EEG signals, identifying key temporal intervals (first and last intervals) that significantly contribute to emotion classification accuracy, enabling more efficient and precise emotion inference.
\end{itemize}

\section{Related work}
\label{sec:relatedwork}
Human emotions can be classified in two main ways \cite{Wenming15}: the discrete basic emotion approach and the dimensional approach. The discrete basic emotion approach categorizes emotions into specific states, such as the six basic emotions: joy, sadness, surprise, fear, anger, and disgust \cite{Ekman92}. In contrast, the dimensional approach depicts emotions as continuous entities, often defined by three dimensions (valence, arousal, and dominance) or simply two dimensions (valence and arousal) \cite{Tengfei20}.

In EEG-based emotion recognition, two fundamental components stand out: EEG feature extractors and classifiers \cite{Jenke14}. Former features are typically categorized into single-channel and multi-channel varieties. Statistical metrics, power spectral density (PSD), or differential entropy (DE) are among the commonly employed single-channel features \cite{Shi13}. Conversely, multi-channel approaches aim to capture inter-channel relationships, such as hemispheric asymmetry in PSD and functional connectivity measures \cite{Li19}. They often rely on correlation, coherence, and phase synchronization metrics to estimate brain functional connectivity. In contrast, our proposed model prioritizes dual-channel features and harnesses a recurrent neural network architecture to integrate them effectively.

In this regard, EEG classifiers can be categorized into two main types: topology-invariant and topology-aware \cite{Xueqi23}. The former, such as Support Vector Machines (SVM) or Recurrent Neural Networks (RNN), do not consider the topological structure of EEG features during learning \cite{Zhong22}. In contrast, topology-aware classifiers like Convolutional Neural Networks (CNN) and Graph Neural Networks (GNN) take into account the inter-channel topological relationships \cite{Li17}. They learn EEG representations by aggregating features from neighboring channels using convolutional operations in Euclidean or non-Euclidean space \cite{Zhong22}. However, existing CNNs and GNNs struggle to capture dependencies between distant channels, potentially overlooking crucial emotion-related information \cite{Li17}. More recently, RNNs have been used to learn spatial topological relations by scanning electrodes in vertical and horizontal directions \cite{Yang21}. Nevertheless, these approaches fail to fully utilize the topological structure inherent in EEG channels \cite{Zhong22}. Topologically close channels may seem distant in their scanning sequence. Our model addresses this by partitioning both hemispheres and using heterogeneous branches (non-shared weights). Prior to RNN application, a Conv1D module handles the topological structure issue.

\section{Methodology}
\label{sec:formulation}

The EEG signals recorded from each electrode \(i\) at a particular time \(t\) can be represented as a continuous function of time \(x_i(t)\). In discrete form, the EEG signal can be represented as a sequence of samples \(x_i[n]\), where \(n\) represents the sample index. 

\textbf{Pre-processing steps.} Three pre-processing steps prepare the signal before extracting the frequency-domain representations. 

1) Re-referencing to Mastoid Channels. Re-referencing involves subtracting the average signals from mastoid channels \(A1\) and \(A2\) from each EEG signal. Let \(x_i(t)\) denote the EEG signal from electrode \(i\) and \(x_{A1}(t)\), \(x_{A2}(t)\) denote the signals from mastoid channels \(A1\) and \(A2\) respectively. The re-referenced EEG signal from electrode \(i\) is denoted as \(x_i^{\text{ref}}(t)\):
    
    \[
    x_i^{\text{ref}}(t) = x_i(t) - \frac{{x_{A1}(t) + x_{A2}(t)}}{2}
    \]
    
2) High-pass and Low-pass Filtering. High-pass and low-pass filters are applied to the re-referenced EEG signals to remove unwanted frequency components. Let \(x_i^{\text{ref}}(t)\) denote the re-referenced EEG signal from electrode \(i\) and \(x_i^{\text{filt}}(t)\) denote the filtered signal. The filtered EEG signal from electrode \(i\) is denoted as \(x_i^{\text{filt}}(t)\):
    
    \[
    x_i^{\text{filt}}(t) = \text{HPF}(\text{LPF}(x_i^{\text{ref}}(t)))
    \]
    
3) Filter Delay Removal. Due to the filtering process, a delay is introduced in the signals. We discard the initial samples from each filtered signal to remove this delay. Let \(x_i^{\text{filt}}(t)\) denote the filtered EEG signal from electrode \(i\) and \(x_i^{\text{prep}}(t)\) denote the pre-processed signal after delay removal. The pre-processed EEG signal from electrode \(i\) is denoted as \(x_i^{\text{prep}}(t)\):
    
    \[
    x_i^{\text{prep}}(t) = x_i^{\text{filt}}(t - \text{delay})
    \]

After pre-processing, the EEG signals are denoted as \(x_i^{\text{prep}}(t)\). The frequency-domain representation obtained through FFT is denoted as \(\phi_i(f)\).

\textbf{Frequency-Domain Representations.} Let's denote this representation as \(\phi_i(f)\), where \(f\) represents frequency.

\[
\phi_i(f) = \text{FFT}\{x_i^{\text{prep}}(t)\}
\]

This represents the Fourier Transform of the EEG signal \(x_i^{\text{prep}}(t)\). The magnitude of  \(\phi_i(f)\) provides information about the amplitude of different frequency components present in the EEG signal, while the phase of  \(\phi_i(f)\) provides information about the timing/delay relationships between these frequency components. These representations allow for EEG data analysis in both the time and frequency domains, enabling insights into brain activity patterns and their associations with cognitive processes such as emotion. Subsequently, the resulting data is segregated into left and right hemispheres for further examination, represented as $\phi_i(f) = \phi_i^{\text{left}}(f) \cup \phi_i^{\text{right}}(f)$. EEG data captured from electrodes situated on the left hemisphere, including $Fp1$, $F7$, $C3$, $P3$, $O1$, $F3$, $T3$, $T5$, $Fz$, $Cz$, and $A1$, is categorized under the left hemisphere, $\phi_i^{\text{left}}(f)$. Conversely, data obtained from electrodes located on the right hemisphere, comprising $Fp2$, $F8$, $C4$, $P4$, $O2$, $F4$, $T4$, $T6$, $Fz$, $Cz$, and $A2$, is attributed to the right hemisphere, $\phi_i^{\text{right}}(f)$.

\textbf{Classifier}. The architecture comprises two parallel streams, each processing EEG signals from different hemispheres. Each stream starts with a 1D convolutional layer followed by max-pooling and dropout regularization, with a dropout rate of $0.5$. The output is reshaped and fed into a Long Short-Term Memory (LSTM) layer to capture temporal dependencies. Another dropout layer is applied to prevent overfitting before flattening the output. This process is repeated for signals from both hemispheres. The processed outputs are concatenated, and the concatenated representation is passed through a fully connected layer with ReLU activation and L2 regularization. Finally, a softmax layer produces probabilities for the six emotion classes. The model is compiled with categorical cross-entropy loss and optimized using the Adam optimizer, with accuracy as the evaluation metric. This architecture allows the model to effectively capture temporal dynamics from EEG signals in order to chose between the six possible emotions. The loss function for the provided code can be expressed as the categorical cross-entropy loss, commonly used for multi-class classification tasks like emotion classification. Mathematically, the categorical cross-entropy loss $L$ can be defined as:

\[
L = -\frac{1}{N} \sum_{n=1}^{N} \sum_{c=1}^{C} y_{n,c} \log(\hat{y}_{n,c})
\]

Where:

\begin{itemize}
    \item $N$ is the total number of samples,
    \item $C$ is the number of classes (in this case, 6 for the six emotions),
    \item $y_{n,c}$ is the true label (one-hot encoded) of sample $n$ for class $c$,
    \item $\hat{y}_{n,c}$ is the predicted probability of sample $\phi_n(f)$ belonging to class $c$ as output by the model.
\end{itemize}

\begin{table}[ht!]
\caption{Accuracy on the validation set for each considered approach.}
\label{table_red}
\centering
\begin{tabular}{|l||c|c|c|} 
\hline
Approach & $\text{Competition}_{\text{base}}$ & Mono-Hemispheric & Bi-Hemispheric\\ 
\hline\hline
Accuracy & $19.4$\% & $23.3$\% & \bm{$28.9$\%} \\
\hline
\end{tabular}
\end{table}

\section{Experimental Setup}
\textbf{Dataset}.
The dataset consists of three sets: training, validation, and test, each serving specific purposes. The training set comprises data from 20 subjects, while the validation set involves 10 subjects. Only the training and validation sets are labeled, with the test set reserved for contest organizers. It contains 720 rows of EEG signal recordings from 21 sensors positioned by the 10-20 International System. Subjects express six fundamental emotions after a neutral state. Each subject repeated each sample three times, with recordings lasting approximately 15 seconds at a resolution of about 300~Hz. The data is stored in two CSV files and has undergone pre-processing, including high-pass and low-pass filtering with cutoff frequencies of 1~Hz and 50~Hz, respectively, a filter delay of 40 ms, and a main frequency of 50~Hz. The sensor data units are measured in microvolts, with the reference location set at the Pz electrode.


\begin{figure}[t]  
    \centering
    \includegraphics[scale=0.4]{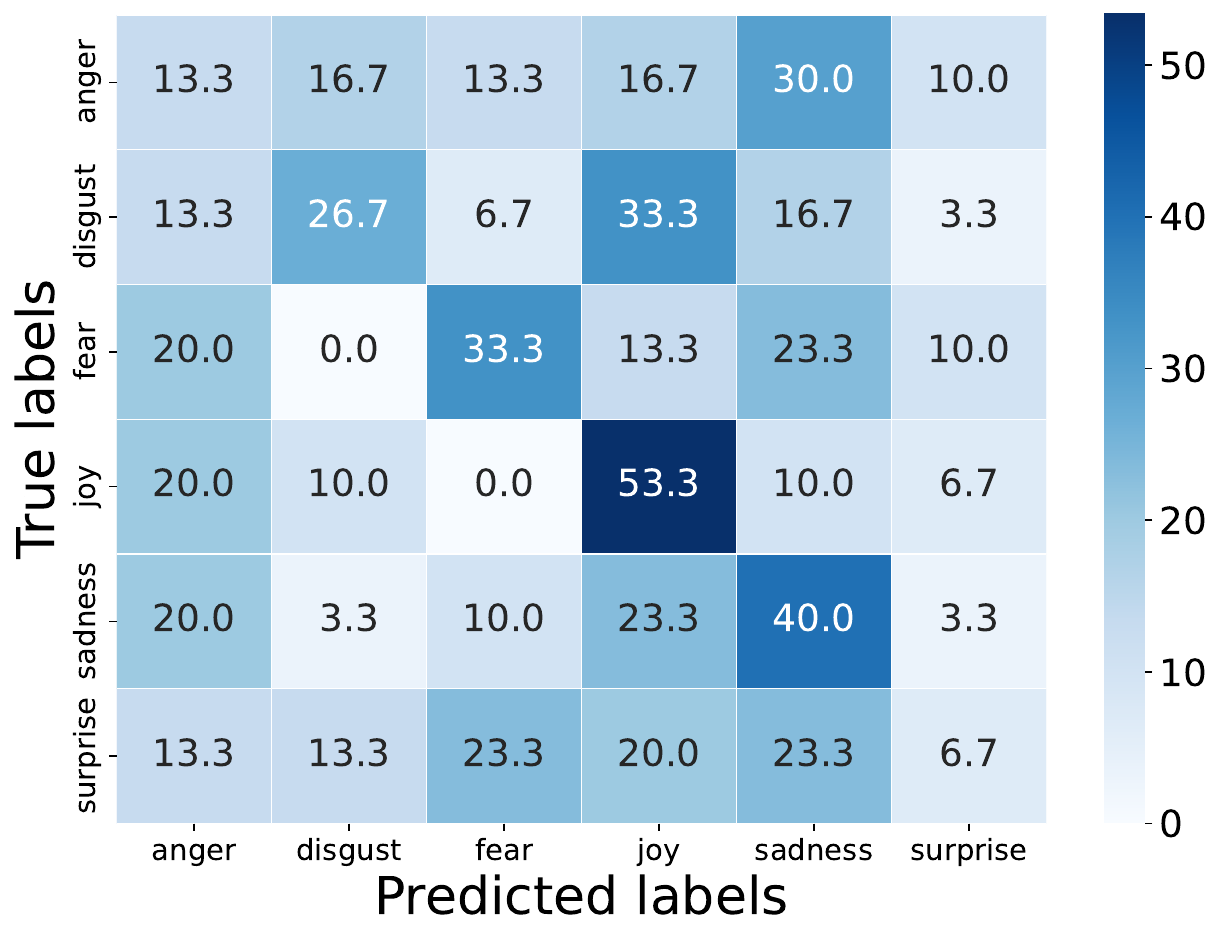}
    \caption{Confusion matrix for the dual-branch approach on the validation set.}
    \label{confusion_matrix}
\end{figure}

\section{Experimental Evaluation}
\label{sec:experiments}

The accuracy results on the validation set for each considered approach are presented in Table \ref{table_red}. The baseline accuracy provided by the competition organizers, denoted as $\text{Competition}_{\text{base}}$, is recorded at $19.4\%$. Subsequent enhancements are observed with the Single Branch approach achieving $23.3\%$, while the Dual Branch model demonstrates the highest accuracy at \textbf{$28.9\%$}. These findings indicate the efficacy of employing a dual-branch architecture for improving emotion classification accuracy compared to both the baseline and single-branch approaches.

When evaluating the dual-branch approach on the validation set, the confusion matrix provides valuable insights into the model's performance across different emotion classes. The confusion matrix, depicted in Figure \ref{confusion_matrix}, illustrates the model's predictions compared to the ground truth labels for six emotions: anger, disgust, fear, joy, sadness, and surprise. Each cell in the matrix represents the percentage of instances predicted as one emotion class (rows) that belong to another class (columns). 

When analyzing the confusion matrix resulting from the Bi-Hemispheric approach on the validation set (see Figure~\ref{confusion_matrix}), it is important to acknowledge the limited number of samples available for training and validation, with only 20 subjects in the training set and 10 subjects in the validation set, as detailed earlier. Despite the small sample size, the matrix reveals notable strengths in the model's performance across various emotion categories. Notably, the model demonstrates commendable accuracy in identifying joy and sadness, with $53.3\%$ and $40.0\%$ of samples correctly classified, respectively. Additionally, the model exhibits promising capability in distinguishing between emotions such as disgust and surprise, achieving $26.7\%$ and $33.3\%$ accuracy rates, respectively. While there may be room for improvement, particularly in accurately classifying instances of anger and fear, the Bi-Hemispheric approach shows potential for effectively capturing the underlying patterns in EEG signals associated with different emotion's states, even within the constraints of a small dataset.



\begin{figure}[t]  
    \centering
    \includegraphics[scale=0.6]{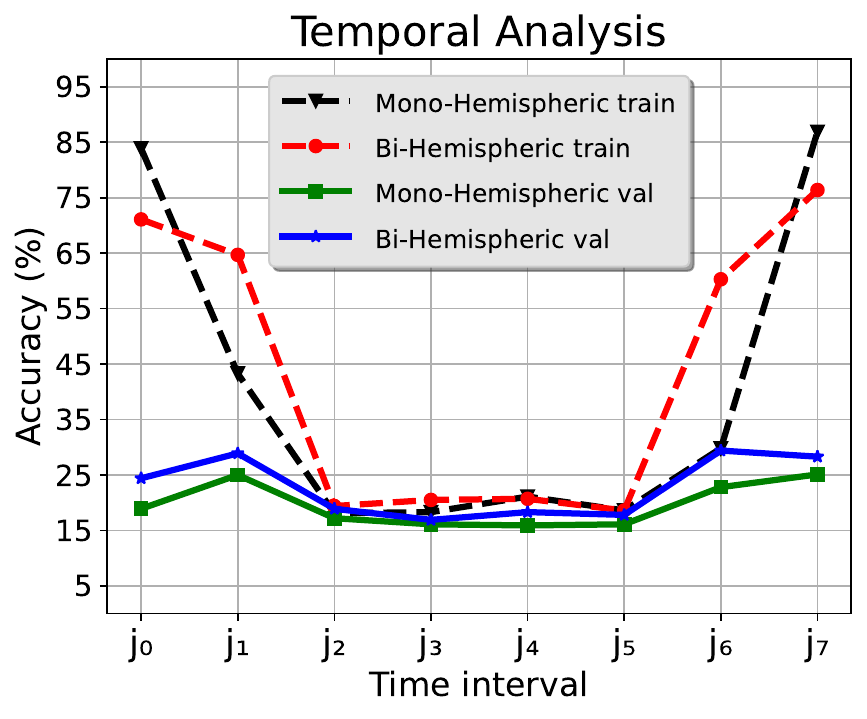}
    \caption{Temporal analysis of training and validation accuracies. Performance of Mono-Hemispheric and Bi-Hemispheric approaches across eight intervals (j0 to j7)}
    \label{tempana}
\end{figure}

\section{Temporal Analysis}
\label{sec:ana}

When performing a temporal analysis, the EEG inputs were divided into eight intervals to classify each interval separately. This temporal segmentation allows for a more granular examination of the EEG signals, potentially capturing transient changes in brain activity over time. Formally, the division of the EEG signal into eight intervals can be expressed as follows:
\[
\text{Interval}_j = \left[ \frac{j \times \text{total\_time}}{8}, \frac{(j+1) \times \text{total\_time}}{8} \right]
\]
Where $j$ represents the interval index ranging from 0 to 7, and $\text{total\_time}$ denotes the duration of the EEG signal.

Analyzing the training and validation accuracies of the Mono-Hemispheric and Bi-Hemispheric approaches across eight intervals (j0 to j7), as depicted in Figure \ref{tempana}, reveals intriguing patterns in the models' temporal performance. Notably, both approaches exhibit higher accuracy at the initial and final intervals (j0 and j7), indicating robust learning at the beginning and end of the temporal sequence. Intervals j1 and j6 notably stand out with notable validation accuracy for both approaches. In particular, interval j1 demonstrates the highest validation accuracy, highlighting the significance of early temporal dynamics in accurately discerning emotion's states. For the Mono-Hemispheric approach, j1 achieves a validation accuracy of 25.0\%, while for the Bi-Hemispheric approach, it reaches an even higher 28.9\%. Similarly, interval j6 also exhibits promising validation accuracy, with the Bi-Hemispheric approach achieving 29.4\% accuracy. However, a noticeable decline in accuracy is observed in the middle intervals (j2 to j5), suggesting challenges in capturing the nuanced changes in EEG signals associated with varying emotion's states during these periods. While the Mono-Hemispheric approach shows a more pronounced decline in accuracy in the middle intervals, the Bi-Hemispheric approach demonstrates relatively better performance, albeit with some fluctuations. This discrepancy highlights the effectiveness of leveraging information from both hemispheres in capturing temporal dynamics, particularly during challenging transition phases between emotion's states.
\section{CONCLUSIONS}
\label{sec:con}

In conclusion, our study addresses the pertinent challenge of emotion classification through EEG signals within the context of the Mapping Human Emotions through EEG Signals FG24 competition. Our proposed two-stream neural network demonstrated superior performance to baseline methods, enhancing emotion recognition accuracy when utilizing a Bi-Hemispheric approach. Furthermore, our temporal analysis revealed that specific signal intervals at the beginning and end of the emotion stimulus sequence significantly contribute to improved accuracy. Our approach effectively captures subtle variations in emotion's states, considering these insights. Notably, in the test set of the competition, where labels are not publicly disclosed, our Bi-Hemispheric approach achieved an average accuracy of 22.78\%, outperforming the baseline method, which achieved an average accuracy of 18.89\%. Specifically, our approach exhibited improved accuracy in classifying joy and sadness emotions, indicating its efficacy in discerning nuanced emotion's states.
Our proposed approach demonstrates robustness and potential applicability in real-world scenarios, promising advancements in EEG-based emotion recognition.




{\small
\bibliographystyle{ieee}

\begin{thebibliography}{10}\itemsep=-1pt

\bibitem{Britton06}
J.~C. Britton, K.~L. Phan, S.~F. Taylor, R.~C. Welsh, K.~C. Berridge, and I.~Liberzon.
\newblock Neural correlates of social and nonsocial emotions: An fmri study.
\newblock {\em NeuroImage}, 31:397--409, 2006.

\bibitem{Ekman92}
P.~Ekman.
\newblock Are there basic emotions?
\newblock {\em Psychological review}, 99 3:550--3, 1992.

\bibitem{Freire23}
D.~Freire-Obreg{\'o}n, D.~Hern{\'a}ndez-Sosa, O.~J. Santana, J.~Lorenzo-Navarro, and M.~Castrill{\'o}n-Santana.
\newblock Towards facial expression robustness in multi-scale wild environments.
\newblock In {\em Image Analysis and Processing -- ICIAP 2023}, pages 184--195, 2023.

\bibitem{Freire09}
D.~Freire-Obreg{\'o}n, M.~C. Santana, and O.~D{\'e}niz-Su{\'a}rez.
\newblock Smile detection using local binary patterns and support vector machines.
\newblock In {\em International Conference on Computer Vision Theory and Applications}, 2009.

\bibitem{Xueqi23}
X.~Gao, C.~Xu, Y.~Song, J.~Hu, J.~Xiao, and Z.~Meng.
\newblock Node-wise domain adaptation based on transferable attention for recognizing road rage via eeg.
\newblock In {\em ICASSP 2023 - 2023 IEEE International Conference on Acoustics, Speech and Signal Processing (ICASSP)}, pages 1--5, 2023.

\bibitem{Jenke14}
R.~Jenke, A.~Peer, and M.~Buss.
\newblock Feature extraction and selection for emotion recognition from eeg.
\newblock {\em IEEE Transactions on Affective Computing}, 5:327--339, 2014.

\bibitem{Li17}
J.~Li, Z.~Zhang, and H.~He.
\newblock Hierarchical convolutional neural networks for eeg-based emotion recognition.
\newblock {\em Cognitive Computation}, 10:368 -- 380, 2017.

\bibitem{Li19}
P.~Li, H.~Liu, Y.~Si, C.~Li, F.~Li, X.~Zhu, X.~Huang, Y.~Zeng, D.~Yao, Y.~Zhang, and P.~Xu.
\newblock Eeg based emotion recognition by combining functional connectivity network and local activations.
\newblock {\em IEEE Transactions on Biomedical Engineering}, 66:2869--2881, 2019.

\bibitem{Yang21}
Y.~Li, L.~Wang, W.~Zheng, Y.~Zong, L.~Qi, Z.~Cui, T.~Zhang, and T.~Song.
\newblock A novel bi-hemispheric discrepancy model for eeg emotion recognition.
\newblock {\em IEEE Transactions on Cognitive and Developmental Systems}, 13(2):354--367, 2021.

\bibitem{Mirsamadi17}
S.~Mirsamadi, E.~Barsoum, and C.~Zhang.
\newblock Automatic speech emotion recognition using recurrent neural networks with local attention.
\newblock In {\em 2017 IEEE International Conference on Acoustics, Speech and Signal Processing (ICASSP)}, pages 2227--2231, 2017.

\bibitem{Pantic03}
M.~Pantic and L.~J.~M. Rothkrantz.
\newblock Toward an affect-sensitive multimodal human-computer interaction.
\newblock {\em Proc. IEEE}, 91:1370--1390, 2003.

\bibitem{Santana22}
O.~J. Santana, D.~Freire-Obreg{\'o}n, D.~Hern{\'a}ndez-Sosa, J.~Lorenzo-Navarro, E.~S{\'a}nchez-Nielsen, and M.~Castrill{\'o}n-Santana.
\newblock Facial expression analysis in a wild sporting environment.
\newblock {\em Multimedia Tools and Applications}, 82:11395 -- 11415, 2022.

\bibitem{Shi13}
L.-C. Shi, Y.~Jiao, and B.-L. Lu.
\newblock Differential entropy feature for eeg-based vigilance estimation.
\newblock {\em 2013 35th Annual International Conference of the IEEE Engineering in Medicine and Biology Society (EMBC)}, pages 6627--6630, 2013.

\bibitem{Tengfei20}
T.~Song, W.~Zheng, P.~Song, and Z.~Cui.
\newblock Eeg emotion recognition using dynamical graph convolutional neural networks.
\newblock {\em IEEE Transactions on Affective Computing}, 11(3):532--541, 2020.

\bibitem{Zheng17}
W.~Zheng.
\newblock Multichannel eeg-based emotion recognition via group sparse canonical correlation analysis.
\newblock {\em IEEE Transactions on Cognitive and Developmental Systems}, 9:281--290, 2017.

\bibitem{Wenming15}
W.~Zheng, H.~Tang, and T.~S. Huang.
\newblock {\em Emotion Recognition from Non-Frontal Facial Images}, chapter~8, pages 183--213.
\newblock John Wiley and Sons, Ltd, 2015.

\bibitem{Zhong22}
P.~Zhong, D.~Wang, and C.~Miao.
\newblock Eeg-based emotion recognition using regularized graph neural networks.
\newblock {\em IEEE Transactions on Affective Computing}, 13(3):1290--1301, 2022.

\end{thebibliography}

}

\end{document}